\documentclass[11pt]{article}
\usepackage{coling2020}
\usepackage{times}
\usepackage{url}
\usepackage{latexsym}

\usepackage{graphicx}
\usepackage{tabularx}
\usepackage{soul}
\usepackage{varioref}
\usepackage{hyperref}
\usepackage{rotating}
\usepackage[table]{xcolor}
\usepackage{epstopdf}
\usepackage[latin1]{inputenc}
\usepackage{xstring}

\title{Data Readiness for Natural Language Processing}

\author{Fredrik Olsson \\
  RISE \\ 
  Sweden \\
  {\tt fredrik.olsson@ri.se} 
  \And
  Magnus Sahlgren \\
  RISE \\ 
  Sweden \\
  {\tt magnus.sahlgren@ri.se} 
  \\}

\date{}

\begin{document}
\maketitle
\begin{abstract}
This document concerns data readiness in the context of machine learning and Natural Language Processing. It describes how an organization may proceed to identify, make available, validate, and prepare data to facilitate automated analysis methods. The contents of the document is based on the practical challenges and frequently asked questions we have encountered in our work as an applied research institute with helping organizations and companies, both in the public and private sectors, to use data in their business processes.
\end{abstract}

\section{Introduction}
\label{sec:intro}

At the Research Institutes of Sweden (RISE)\footnote{\url{https://ri.se}}, we work in cooperation with other organizations and companies, both in the public and private sectors,
with research and innovation in Natural Language Processing (NLP).  A major challenge that we often encounter is a lack of readiness with respect to data. Even if the research problem is sufficiently well defined, and the business value of the proposed solution is well described, it is often not clear what type of data is required, if it is available, or if it at all exists. We find that there is often not even a framework available to discuss issues related to data. The purpose of this document is to outline and highlight issues related to data accessibility, validity, and utility that may arise in such situations. We hope that this document may serve as a guide for working practically with data in the context of applied NLP.

\subsection{Scope}

This document is concerned exclusively with data readiness in the context of NLP. Other modalities such as images, video, or sensor data are not covered, but similar considerations apply in those cases.

Work on data readiness related to other forms of data include that of \newcite{nazabal2020data}, who address data wrangling issues from a general stand-point using a set of case studies, as well as the work by \newcite{van2019quality}, and \newcite{harvey2019standardised} that both deal with data quality in medical imaging. We have not found any work that focuses specifically on data readiness in the context of NLP.

\subsection{How to use this document}

The intention is for this document to provide insights into the type of challenges one might encounter, with respect to data, when embarking on a project involving NLP. The document is focused on asking the right questions rather than providing an explicit guide that covers all possible challenges in a project: such a guiding will inevitably vary with the specific task at hand. The following four sections make up the document:

\begin{itemize}
    \item {\em Data Readiness Levels} introduce the notion of data readiness.
    \item {\em Project phases} outlines the typical structure of a research or innovation project, and puts data readiness into context within that structure.
    \item {\em Questions for guidance} is the central part of the document. Use the questions to understand the data readiness of your own project.
    \item {\em How to contribute} contains instructions for how to help us improve the document. Please, reach out with any questions, or suggestions!
\end{itemize}

\section{Data Readiness Levels}

The notion of Data Readiness Levels (DRLs) introduced by \newcite{lawrence2017data}, provides a way of 
talking about data much in the same way Technology Readiness Levels facilitate  
communication regarding the maturity of technology \cite{NasaTrl2017}. As such, it is a framework suitable for
exchanging information with stakeholders regarding data {\bf accessibility}, {\bf validity}, and {\bf utility}.

Figure \ref{fig:drl-table} below illustrates the three different major Bands of the DRLs. Each band can be thought of as consisting of multiple
levels. At the lowest level, i.e., Band C - Level C-4, someone has heard that there is data to be had. This "hearsay" level is
often where a new project starts --- someone knows that there should be data available to work with. Since the access
to data often dictates the bounds of possible analysis, and thus the level of results attainable, walking a new project
or stakeholder through the DRLs is not only a nice-to-have, it is a must-do. If there is no data to work with, it does
not matter what kind of algorithms are available.

\vspace{1em}

\begin{figure*}[!ht]
\begin{center}
\includegraphics[scale=0.17]{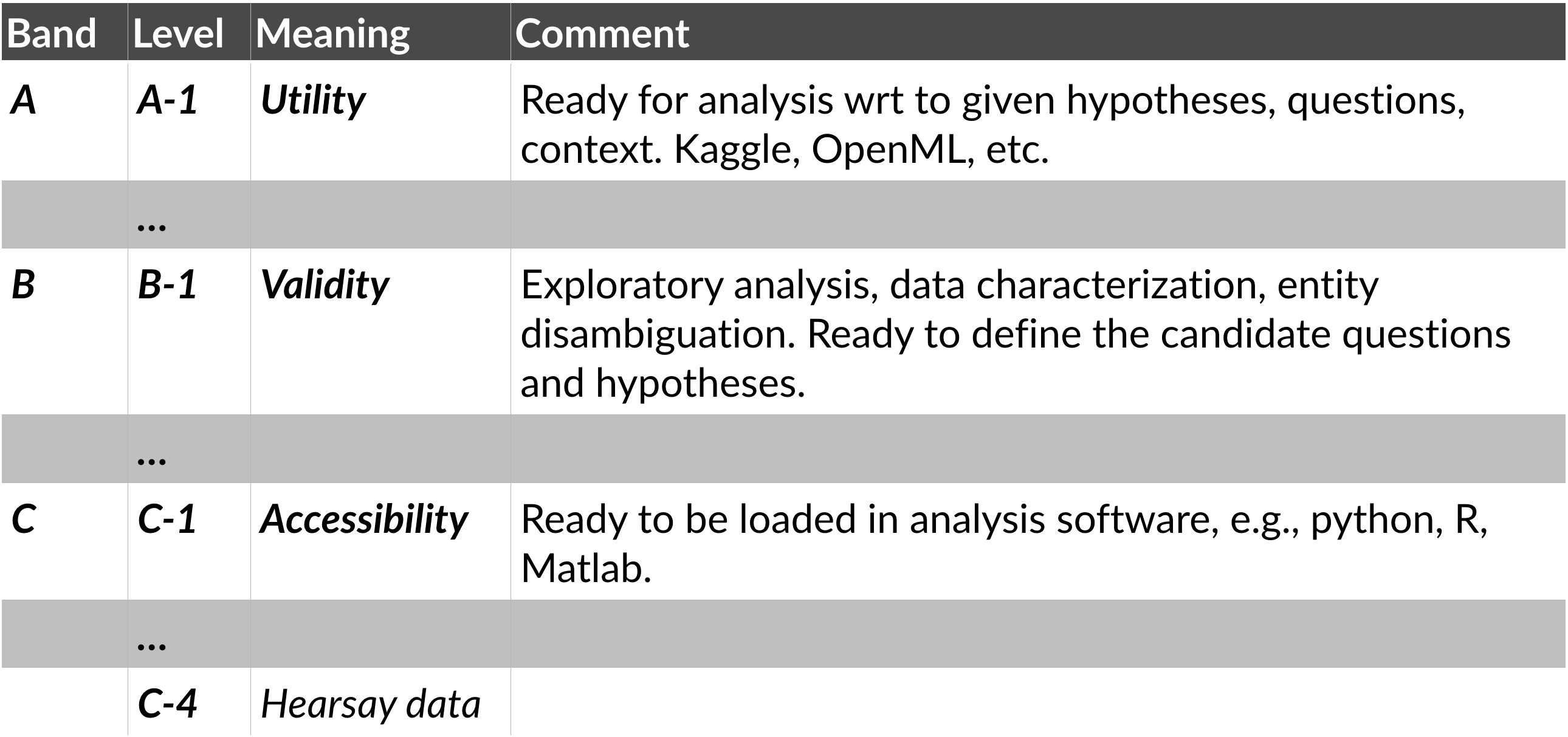}
\caption{An overview of the different bands of Data Readiness Levels.}
\label{fig:drl-table}
\end{center}
\end{figure*}

\subsection{Band C}

{\bf Band C} concerns the {\bf accessibility} of data. All work at this level serves to grant the team 
intended to work with the data access to it; once access is provided, the data is considered to be at Band C - Level C-1, and ready to be brought into Band B. Issues that fall under Band C include:
\begin{itemize}
    \item {\bf Does the data exist?} Is the data required to address the task even recorded?
    \item {\bf Data conversion and encoding}. One of the major challenges faced within NLP is the conversion of
documents from a source format, e.g., PDF, Word or Excel, to a format suitable for addressing the task at hand. 
In order to move beyond Band C, data conversion and encoding have to be in place.
    \item {\bf Legal aspects of accessibility}. Not only should the data be available to the intended team, but the team and the
  result of their efforts to produce a solution to the task at hand should also be cleared with respect to the legal aspects
  of accessing, and handling of the data. This include, e.g., the handling of personal identifiable information, and copyright issues.
    \item {\bf Programmatic aspects of accessibility}. The team should have easy access to the data, by a method of their choice, 
  e.g., via an API, or a database interface.

\end{itemize}

\subsection{Band B}

{\bf Band B} concerns the {\bf validity} of data. In order to pass Band B, the data has to be valid in the sense that it is
representative of the task at hand. Furthermore, the data should be deduplicated, noise should be identified, 
missing values should be characterized, etc. At the top level of Band B, the data should be suitable for exploratory analysis, and the 
forming of working hypotheses.

\subsection{Band A}

{\bf Band A} concerns the {\bf utility} of data. The utility of the data concerns the way in which the data is intended to be
used: Is the data enough to solve the task at hand?

A project should strive for data readiness at Band A - Level A-1. Note that the Data Readiness Levels should be interpreted with respect to a given task. As an example, data that is at Band A for the task of pre-training a language model might be considered Band B for the task of training a Named Entity Recognizer, simply because the requirements on the data are different in the two tasks. For training a language model, it could be enough if the data is unlabelled, clean and drawn from domains of general interest, while for the Named Entity Recognizer, the data will need to be annotated with the appropriate information, and taken from the very specific domain in which the recognizer will then operate.

\section{Project phases}

To put the need for training, validation, and evaluation data into context, let's have a look at the structure of a typical NLP project. There are many variants of the way organizations work with applied research, innovation,
and analytics. Most of them progress roughly by the following steps:

\begin{enumerate}
    \item Specify the problem to solve.
    \item Assess the data available to produce and evaluate solutions to the problem.
    \item Select technology to use for solving the problem.
    \item Implement a solution on the form of a demonstrator, prototype or product.
    \item Mutual transfer of knowledge between the research team and the client.

\end{enumerate}

The above steps are usually carried out in an iterative fashion in that, e.g., the problem specification is re-visited and updated as the knowledge of the data available is increased, the transfer of knowledge is omnipresent in all stages of the project, and the technology choices and implementation are iteratively updated and tested as the project progresses.

Although the Data Readiness Levels described in the previous section permeate all stages of the project management, steps 1 through 3 in the project structure above are where the data readiness is usually addressed in depth. Thus, it is crucial
that the Data Readiness Levels are in order at the beginning of the project.

Common data science, and analytics project processes include 
{\em Cross-industry standard process for data mining} (CRISP-DM, \newcite{Shearer2000}), and {\em The Team Data Science Process} 
(TDSP, \newcite{MSFT-TDSP2020}).

\section{Questions for guidance}

The questions below are intended to serve as guidance in the process of reaching the appropriate Data Readiness Levels for solving
a research or business problem related to NLP.

\subsection{What problem are you trying to solve?}

The first step in every applied project is to make sure that there is a clear and concise definition of done; what is the goal of the project, and how do we know when we have reached it? We recommend that the goal is intimately tied to a tangible business value, and that there is a clear idea already from the start how the project fits into the business value chain. To put the point succinctly: if the project is not motivated by a clear business need, it will not deliver any business value. When specifying the problem, make sure to focus on the actual business need that is to be fulfilled, as opposed to 
a technology to solve the problem. Try to make all assumptions about the business need explicit; write them down, and
make sure to vet the assumptions thoroughly with the stakeholders to have them sign-off on a common understanding of
the problem. As with everything else, you should expect to iterate on the specification of the problem, and as the process proceeds,
the adjustments to the specification will become smaller.

Having a detailed and agreed-on specification of the business need and problem is crucial since seemingly subtle details
may have a large impact on the space of possible techniques to actually solve the problem. For instance, if the 
stakeholder is expecting an implemented solution to extract, rank, and present passages from long documents 
with sub-second latency to a user, from a data point-of-view you have to know whether the document set is expected to 
change continuously, what format the documents are in, and what information there are to guide the learning-to-rank
process. 
The characteristics of the problem you set out to solve dictates the requirements on the data needed to solve it.
Thus, the specification of the problem is crucial for understanding the requirements on the data in terms of, e.g., 
training data, and the need for manual labelling of evaluation or validation data. Only when you know the characteristics of the data, it will be possible to come up with a candidate 
technological approach to solve the problem.

\subsection{What data are available to you?}

What does "available to you" mean? There is a difference between knowing that there is data to be had, and actually having
the appropriate access to the right data at the right time. The Data Readiness Levels framework provides a good 
way of talking to stakeholders, both data owners and business problem owners. Being "available" in this context, means
that the data is at Band B or above.

With data at Band B, you will most likely be able to form working hypotheses about the suitability of the data with
respect to the problem you are trying to solve. For example to gauge whether the data contains the information you need;
the amount of data is sufficient; or there are any legal constraints for using the data.

\subsection{Are you allowed to use the data available?}

The legal aspects of using the data depend on many things. As such, the legalities should be considered a primary citizen
of the Data Readiness Level assessment with respect to your particular challenge. Make sure you involve the appropriate 
legal competence early on in your project. Matters regarding, e.g., personal identifiable information, and 
GDPR have to be handled correctly. Failing to do so may result in a project failure, even though all technical aspects 
of the project are perfectly sound.

\subsection{What data do you need to solve the problem?}

Given the insight into what data is available, ask yourself the questions: 

\begin{itemize}
    \item What data do you need to solve the problem?
    \item Is that a subset of the data that is already available to you?
    \item If not: is there a way of getting all the data you need?
\end{itemize}

If there is a discrepancy between the data available, and the data required to solve the problem, that discrepancy
has to be mitigated. If it is not possible to align the data available with what is needed, then this is a cue to
go back to the drawing board and either iterate on the problem specification, or collect suitable data. 
Perhaps there are other ways to solve the business or research need than what has been specified so far?
It is not uncommon to discover that the data actually needed to solve a problem does not even exist, despite initial assumptions to the contrary. Such a conclusion need not be entirely negative, since it opens up the possibility to initialize a data collection effort, or, if needed, to refine business processes to start generating relevant data.

\subsection{How do you know if you have succeeded in solving the problem?}

When you are in the process of defining and specifying the problem to solve, you should also consider how to {\em evaluate}
the potential solutions to the problem. 

The type of data required to evaluate a solution is often tightly connected to
the way the solution is implemented: if the solution is based on supervised machine learning, i.e., requiring labelled examples, 
then the evaluation of the solution will also require labelled data. For example,
if the problem you face is to help users to find topically relevant sections in a large number of yearly reports
submitted by public agencies, then you will most likely need to construct a collection of sections labelled with the 
appropriate topic descriptors. Such data can then be used to assess and compare the technical solutions you come up with. 

On the other hand, if it is possible to produce a
solution based on unsupervised machine learning, then {\em perhaps} it is possible to conduct the evaluation based on 
unlabelled data too (although it is far from certain).

In any case, if the solution depends on labelled training data, the process of annotation usually also results in the 
appropriate evaluation data.
Any annotation effort should take into account:

\begin{itemize}
    \item {\bf The quality of the annotations}. The agreement between the annotators working on the data, aka the inter-annotator
 agreement, provides a good starting point for assessing the quality of the annotations overall. The reasons 
 for low inter-annotator agreement can be related to, e.g., 
 unclear annotation guidelines, difference in expertise among the annotators, that the task is simply too hard, or a 
 combination of all of the above. The annotations produced should be carefully monitored with respect their quality. 
 Deviations in quality over time should be analyzed so as to facilitate mitigation of a potential decrease in the capabilities of the
  model that relies on the annotated data for training. 
    \item {\bf Temporal aspects of the data characteristics}. How often do the distribution of the data to learn from change, i.e.,
how often do we need to produce new annotations? When do we know that we need newly annotated data?
    \item {\bf Representativity of the data}. Is the data annotated really suitable for the task at hand? Does it reflect the way
users interact with the system?

\end{itemize}

Apart from the more quantitative approach to evaluation facilitated by the use of annotated test data, the qualitative aspects of evaluation should also be considered. Qualitative evaluation can take on the form of user acceptance tests, and surveys for identifying, e.g., nonsensical output from a model, or to get a read on the reduction of cognitive load that a user of a system experiences. Furthermore, qualitative evaluation might help identifying issues with missing data that, if it present, would help increase the performance of the model.

Obtaining the training, evaluation, and validation data is at the core of producing a machine learning-based solution to
a problem. The quality of the data sets the upper bound to what can be achieved by the learned functionality. Also included
in the production process are issues such as model selection, setting up infrastructure for machine learning, continuously
monitoring the solution's performance for decrease in performance, etc. A good overview of the end-to-end process is presented  by \newcite{Ameisen2020}.

\subsection{How does your organization store new data?}

Even if the data processing in your organization is not perfect with respect to the 
requirements of machine learning, each project you pursue has the opportunity to articulate improvements to your
organization's data storage processes. Ask yourself the questions: How does my organization store incoming data? Is
that process a good fit for automatic processing of the data in the context of an NLP project, that is, is the
data stored on a format that brings it beyond Band C ({\bf accessibility}) of the Data Readiness Levels? If not; what
changes would need to be made to make the storage better?

A couple of things we have found important over the course of multiple projects are related to the format in which
the data is stored. In particular:

\begin{itemize}
    \item {\bf Information in the data should not be removed prior to storing it}. Destructive processing, such as tokenization, 
stemming, and downcasing of text should not be carried out as a part of the data storing process. Do not conflate the intended usage of the data
for a particular use case with the storage format; make as few assumptions about how the data will be used in the 
future as possible.
    \item {\bf Avoid proprietary formats, and formats not intended for automatic processing}. Document formats output by regular 
word processing software, for instance PDF, Word, and Pages, are {\em not} appropriate formats for input to automatic, machine learning-based
processing of information. The challenges of converting, e.g., a PDF file to a textual format suitable for use in a processing
pipeline are many. There are currently no general and stable programmatic solution for avoiding:
\begin{itemize}
    \item the omission of information, e.g., erroneously broken up words; 
    \item the introduction of superfluous information, e.g., insertion of page headers as part of the text;
    \item character encoding issues;
    \item the dispersion of information from tables into running text; 
    \item the mix up of the order of paragraphs or columns, or;
    \item a number of additional challenges, as described by \newcite{Panait2020}.
\end{itemize}
    \item {\bf Logical structure}. If possible, make sure the logical structure of a document is made accessible upon retrieval. 
That is, ensure that relevant parts of a document are possible to refer to by their function, enabling document
segmentation queries such as:
\begin{itemize}
    \item Give me the table of contents of this document.
    \item List, in order, all top level subsections and their textual contents from Chapter 3 in this book.
    \item Extract all tables from this yearly report.
\end{itemize}
\end{itemize}

Failing to address the above issues may result in your data never making it past Data Readiness Level Band C, and
thus not be appropriate for automatic analysis.

\section{How to contribute}

The NLP data readiness document is work-in-progress. If you have any questions, suggestions for edits, or other input, please email the author or submit a pull request in the document's GitHub repository, available at: \url{https://github.com/fredriko/nlp-data-readiness}.

General step-by-step instructions for how to contribute to create a pull request is descibed by, e.g, \newcite{DataSchool2020}.

\section{Acknowledgements}
Comments and input on the text were kindly probided by Fehmi ben Abdesslem, Fredrik Carlsson, Anders Arpteg, and Armin Catovic.

\bibliographystyle{acl}
\bibliography{main}

\end{document}